\documentclass[11pt]{article}
\usepackage[latin1]{inputenc}
\usepackage[english]{babel}
\usepackage[namelimits]{amsmath}
\usepackage{amssymb}
\usepackage{amsmath}
\usepackage{amsthm}
\usepackage{authblk}

\begin{document}
\title{{\bf The spectrum of the Schr\"{o}dinger Hamiltonian for trapped particles in a cylinder with a topological defect perturbed by two attractive delta interactions }}
\author[1,2,3]{S. Fassari\thanks{silvestro.fassari@uva.es}}
\author[2,3]{F. Rinaldi\thanks{f.rinaldi@unimarconi.it}}
\author[2,4,5]{S. Viaggiu\thanks{s.viaggiu@unimarconi.it and viaggiu@axp.mat.uniroma2.it}}
\affil[1]{Departamento de F\'{\i}sica Te\'{o}rica, At\'{o}mica y \'{O}ptica, and IMUVA, Universidad de Valladolid, 47011 Valladolid, Spain}
\affil[2]{Dipartimento di Fisica Nucleare, Subnucleare e delle Radiazioni, Universit\'a degli Studi Guglielmo Marconi, Via Plinio 44, I-00193 Rome, Italy}
\affil[3]{CERFIM, PO Box 1132, Via F. Rusca 1, CH-6601 Locarno, Switzerland}
\affil[4]{Universit\`a di Roma ``Tor Vergata'', Via della Ricerca Scientifica, 1, I-00133 Roma, Italy.}
\affil[5]{INFN, Sezione di Napoli, Complesso Universitario di Monte S. Angelo,
	Via Cintia Edificio 6, 80126 Napoli, Italy.}
\date{\today}\maketitle

\begin{abstract}
In this paper we exploit the technique used in \cite{A}-\cite{5b} to deal with delta interactions
in a rigorous way
in a curved spacetime represented by a cosmic string along the $z$ axis. This mathematical machinery is applied in order to study 
the discrete spectrum of a point-mass particle confined in an infinitely long cylinder with a conical defect on the 
$z$ axis and perturbed by two identical attractive delta interactions symmetrically situated around the origin. We derive a suitable approximate formula for the total energy. As a consequence, we 
found the existence of a mixing of states with positive or zero energy with  the ones with negative energy (bound states). 
This mixture depends on the radius $R$ of the trapping cylinder. The number of quantum bound states is an increasing function
of the radius $R$. It is also interesting to note the presence of states with zero total energy (quasi free states). Apart from
the gravitational background,
the model presented in this paper is of interest in the context of nanophysics and graphene modeling. In particular, the graphene
with double layer in this framework, with the double layer given by the aforementioned delta interactions and the string on the $z-$axis modeling topological defects connecting the two layers.
As a consequence of these setups, we obtain the usual mixture of positive and negative bound states present in the graphene literature.
\end{abstract}

\section{Introduction}
Since the final formulation of general relativity, a great deal of theoretical work (see for example \cite{6} and references therein)
has been done in order to unify general relativity and quantum mechanics (loop string, causal sets, non-commutative geometry...).
Unfortunately, a shared quantum gravity theory is still lacking. Technical difficulties 
are related to the covariance of general relativity under coordinate transformations and to the extreme difficulty to define localized energies in general relativity.\\
In practice, the main issue is due to the lack of an experimental signature as a guidance for theoretical physicists.
To this purpose, it is important to study possible detectable effects induced by gravity on given quantum systems. As an example, the effects induced by some specific general
relativistic background on a hydrogen atom has been investigated in \cite{7,8}, where a shift effect on the spectrum arisises depending on the particular chosen background. Similar shift effects can also be found in \cite{9,10}. Moreover, the experiments performed in
\cite{11} confirmed the effects induced by a gravitational field on the phase difference between two neutron beams.
Another phenomenon is due to the interaction gravity-quantum system,
for instance neutrino oscillations \cite{12}.
Another intriguing line of research is related to the well-known Aharonov-Bohm effect \cite{13}
where a charged particle can interact
with an electromagnetic field also in those
regions with vanishing field. In general relativity similar phenomena are well known
in presence of topological defects generated by a cosmic string \cite{14}. In this framework, 
the metric outside the string is
locally flat and, as a consequence, a particle at a fixed position is not influenced by a gravitational field. 
Nevertheless, the topology generated by a string along the $z$-axis is not the one of a Minkowskian spacetime, but rather it 
shows a conical defect.
This defect produces many effects on traveling test particles, for example, gravitational lensing \cite{15}, pair 
production \cite{16} and the gravitational counterpart of the Aharonov-Bohm effect \cite{17}.\\
The study of the effects arising from an infinitely long cosmic string on a non-relativistic quantum system has been further analysed in \cite{18}, where, shift effects given by a cosmic string arise in particular
by considering a quantum system with harmonic and Coulombian potential.
Inspired by the investigation carried out in \cite{18}, we consider a quantum system perturbed by two identical attractive delta interactions symmetrically situated around the origin.
As shown in \cite{A}-\cite{5b}, the spectrum of the Schr\"{o}dinger Hamiltonian perturbed by spatially symmetric point interactions can
provide interesting shift effects after using the suitable mathematical treatment. \\
It is thus interesting, in line of the reasonings above, to investigate the possible physical consequences of quantum shift effects present in the papers \cite{A}-\cite{5b} in a curved background and in particular in the presence of topological
(gravitational) defects induced by a cosmic string. This line of research can also be of great interest
in the context of graphene modeling, where the topological defects in a curved background represented by a cosmic string can be used
to depict defects present on the graphene \cite{21,21b,21c}.
The structure of the paper is the following. In section 2 we present the Hamiltonian of the model to be studied. 
Section 3 is devoted to the study of the radial part of the Schr\"{o}dinger equation with Dirichlet boundary conditions.  
In section 4 we exploit the mathematical tools presented in \cite{1} which perfectly agree with the more general findings of section II.2.1 in \cite{B}. In section 5 we study the ground state and first excited energy spectrum formulas obtained in section 4, by restoring the usual unit leghts.
In section 6 we apply the machinery of section 4 and 5 to modelize the double layer graphene with the string, as a topological defect, interconnecting the layers. 
Section 7 is devoted to the conclusions and final remarks.

\section{The model}
To start with, we consider the exterior metric of an infinitely long static string along the $z$ axis. Its energy-momentum tensor
is nothing else but $T_{ab}=\mu\delta(x)\delta(y)diag(-1,0,0,1)$, where we used Cartesian coordinates $(t,x,y,z)$ with metric
signature $(-,+,+,+)$ and $\mu$ is a constant linear mass density of the string. We introduce the deficit angle $B$ with
$B=1-4G\mu/c^2$. In cylindrical coordinates $(t,\rho,\phi,z)$ the line element of a cosmic string is
\begin{equation}
ds^2=-dt^2+d{\rho}^2+B^2{\rho}^2d{\phi}^2+dz^2.
\label{1}
\end{equation}
For positive values of the string density $\mu$ with $B\in[0,1]$, we have the so-called deficit angle. In fact, by performing the
coordinate transformation $\phi\rightarrow B\phi$ we have formally a Minkowskian metric with $\phi$ in the domain
$\phi\in[0,2\pi B]$. For $\mu<0$ we have the so-called surplus-deficit angle metric with $B>1$. Such a situation is exotic in the context of
general relativity since a negative energy density does not satisfy the weak energy condition (i.e. the positivity of the 
energy density for any timelike experimenter). Nevertheless, negative energies can be allowed in a quantum field theory context where fluctuations can induce negative energies (Casimir effect). Strings with negative (attractive) energy density can be considered in 
the framework of graphene modeling (see for example \cite{21}). 

In a static background \cite{18} we can write the Schr\"{o}dinger equation using the Laplace-Beltrami Laplacian
${\nabla}^2_{lb}$. By denoting with $g$ the determinant of the spatial metric, obtained at $t=constant$, with
$g_{ij}$ the spatial metric ($\{i,j\}=\{1,2,3\}$) we have 
${\nabla}^2_{lb}=g^{-\frac{1}{2}}{\partial}_i\left(g^{ij}\sqrt{g}\;{\partial}_j\right)$. 
The Schr\"{o}dinger equation for the wave function
$\psi=\psi(t,x^i)$ of a quantum particle with rest mass $M$ interacting with a static  external potential $V_{ext}=V(x^i)$
becomes:
\begin{equation}
\imath\hbar\frac{\partial\psi}{\partial t}=-\frac{{\hbar}^2}{2M}{\nabla}^2_{lb}\psi+V_{ext}\psi,
\label{2}
\end{equation}
with the Hamiltonian $H$ given by $H=\frac{p^2}{2M}+V_{ext}$. We can now specify the equation (\ref{2}) with the metric (\ref{1}).
For ${\nabla}^2_{lb}$ we have
\begin{equation}
{\nabla}^2_{lb}={\nabla}^2_B=
\frac{{\partial}^2}{\partial{\rho}^2}+\frac{1}{\rho}\frac{\partial}{\partial\rho}+
\frac{{\partial}^2}{\partial{z}^2}+\frac{1}{B^2{\rho}^2}\frac{{\partial}^2}{\partial{\phi}^2}.
\label{3}
\end{equation}
The Laplacian (\ref{3}), can be perturbed by
delta interaction potentials $V_{ext}$.

In order to use the mathematical machinery in \cite{1}, we must
specify the expressions for the external potential $V_{ext}$.
To obtain a discrete spectrum, the L-B operator (\ref{3}) must be perturbed by an attractive potential.
We work in a Lorentzian manifold with $3+1$ dimensions. Hence, the static external potential generally depends on the three
spatial coordinates $\{\rho,z,\phi\}$.\\
We consider only the case of attactive potentials looking like 
$V_{ext}\sim -\lambda\delta(z\pm z_0)$ with $\lambda >0$. 
In this case, the support of the delta is provided by the planes 
$z=\pm z_0$.\\
Due to the static nature of the metric (\ref{2}), we 
can consider the stationary form of (\ref{2}), with  
$H\psi(\rho,z,\phi)=E\psi(\rho,z,\phi)$ and $H=H_0+V_{ext}$, $H_0$ being the free Schr\"{o}dinger Hamiltonian.
After setting the geometrized 
units with $G=c=\hbar=1$, and  by posing, without loss of generality, $M=1/2$
\footnote{The mass choice $M=1/2$ is in agreement with the one present in the paper \cite{1}. Obviously this choice does not represent 
	a loss of generality since we can always take the transformations $E\rightarrow E/2$ and
	$\lambda\rightarrow \lambda/2$ and the usual choice $M=1$ is regained.}, we have
\begin{equation}
-\left[\frac{{\partial}^2}{\partial{\rho}^2}+\frac{1}{\rho}\frac{\partial}{\partial\rho}+
\frac{{\partial}^2}{\partial{z}^2}+\frac{1}{B^2{\rho}^2}\frac{{\partial}^2}{\partial{\phi}^2}\right]\psi
+V_{ext}(\rho,z)\psi=\imath\hbar\frac{\partial}{\partial t}\psi.
\label{4}
\end{equation} 
Let us set
\begin{equation}
\psi(t,\rho,z,\phi)=ke^{\left(-\imath\frac{Et}{\hbar}\right)}\psi(\rho,z,\phi),
\label{5}
\end{equation}
where $k\in{\mathbb{R}}^+$ is the normalization constant.\\
First of all, we consider the following expression for the perturbed potential $V_{ext}$:
\begin{equation}
V_{ext}(\rho,z)=V_{ext}(z)=-\lambda\left[\delta(z-z_0)+\delta(z+z_0)\right],\;\;\lambda>0.
\label{6}
\end{equation}
The potential (\ref{6}) represents an attractive delta perturbation with support on the planes $z=\pm z_0$.

By setting $\psi(\rho,z,\phi)={\psi}(\rho)\xi(z)\zeta(\phi)$ with 
$\zeta(\phi)\sim e^{\imath a B\phi},\;a\in\mathbb{R}$ (see next section for more details), we have
\begin{eqnarray}
& &-\left[\frac{d^2\psi(\rho)}{d{\rho}^2}+
\frac{1}{\rho}\frac{d\psi(\rho)}{d\rho}-\frac{a^2}{{\rho}^2}\psi(\rho)\right]=E_B \psi(\rho), \label{7}\\
& & -\frac{d^2\xi(z)}{dz^2}
-\lambda\left[\delta(z-z_0)+\delta(z+z_0)\right]\psi(z),
=E_z\xi(z),\label{8}
\end{eqnarray}
where
\begin{equation}
E=E_B+E_z.
\label{9}
\end{equation}
It is worth to be noticed that in \cite{1} the following potential has been considered:
\begin{equation}
V_{ext}(\rho,z)=V_{ext}(z)=-\lambda\left[\delta(z-z_0)+\delta(z+z_0)\right]\mp\chi\delta(z),\;\;\lambda>0,
\label{6b}
\end{equation}
where the added delta interaction, $\mp\chi\delta(z)$, represents the attractive and repulsive point interaction placed in the origin.
In this paper, however, we focus our attention on the double layer graphene model. Hence, the above mentioned added interaction is not of interest in our framework, in particular the repulsive case studied in section II.2.1 of \cite{B}.\\ In the general relativity context, the condition $\mu>0$ is required in order to have a non-exotic matter content for the string.
However, the case $\mu<0$ can be of interest, for example, in the modeling of graphene \cite{21}.
The generalized eigenvalues $E_B$, related to the wavefunction $\psi(\rho)$, give rise to a continuous spectrum. We can also consider interesting cases where the 
particle is 'trapped' \footnote{A similar technique has been
used in \cite{23} in order to obtain the spectrum of trapped gravitons by linearization of 
Einstein's equations \cite{23b}.} inside an infinitely long cylinder of radius ${\rho}_0$, so that
we
can introduce the infinite potential wall $V$ with  $V=0$ for $\rho\in(0,{\rho}_0)$ and $V=\infty$ for
${\rho\geq{\rho}_0}$.\\ 
In practice, a 'huge' wall  trapping the particle is introduced and as a 
consequence, the particle is trapped within the cylinder of radius ${\rho}_0$. This modeling could be of interest
to study the behavior of electrons inside a cylindrical conductor with a topological defect along the $z$-axis, and perturbed by two identical attractive delta interactions symmetrically situated around the origin. In this case the Dirichlet boundary condition 
$\psi(\rho={\rho}_0=R)=0$ must be imposed. This boundary condition is suitable, for example, for electrons in a finite metal or in the quantum-confined Stark effect modulator \cite{24}.
Concerning the wave function ${\psi}(\rho,z,\phi)=k\psi(\rho)\xi(z)\zeta(\phi)$, the normalization requires:
\begin{equation}
k^2\int_0^{2\pi}d\phi\int_0^{{\rho}_0}\rho d\rho\int_{-\infty}^{+\infty}{\psi}^2(\rho,z,\phi)dz=1.
\label{10}
\end{equation}
Note that the exact value of $k$ plays no role in result of this paper and we omit the calculation of the integral (\ref{10}).

\section{Radial spectrum of a point-mass particle with a topological defect}
As is well known, the Schr\"{o}dinger equation (\ref{4}) can be separated by using cylindrical coordinates, as shown in (5), (7) and (8). 
The equation with $\psi(z)$ will be analyzed in the next section together with the explicit expressions for the spectrum of the ground state and of the first excited state. Concerning the $\zeta(\phi)$ component, from (\ref{4})-(\ref{5}) we get:
\begin{equation}
\frac{d^2\zeta(\phi)}{d{\phi}^2}+a^2 B^2\zeta(\phi)=0.
\label{11}
\end{equation}
The solution of (\ref{11}) is 
\begin{equation}
\zeta(\phi)\sim e^{\imath aB\phi}
\label{12}
\end{equation}
with $a\in\mathbb{R}$. In order to guarantee the continuity of the wave function, we have to impose
$\zeta(0)=\zeta(2\pi)$. This condition is not trivial in presence of a topological defect ($B\neq 1$). In fact, the 
effective range of the angular part of the spacetime is $\phi\in[0,2\pi B]$ and the surfaces 
$\phi=0$ and $\phi=2\pi B$ must be connected, in order to have a well-defined manifold with metric (\ref{1}). For $B<1$, the range of
$\phi$ is less than $2\pi$, while for $B>1$ we have a surplus angle with $\phi$ greater than $2\pi$.
Then we obtain
\begin{equation}
1=e^{2\pi \imath Ba},
\label{13}
\end{equation}
so that, due to Euler's relation, we get
\begin{equation}
a=\frac{n}{B},\;\;\;n\in\mathbb{Z}.
\label{14}
\end{equation}
From expression (\ref{14}), for the radial equation (\ref{7}), we obtain
\begin{equation}
{\rho}^2\frac{d^2\psi(\rho)}{d{\rho}^2}+\rho\frac{d\psi(\rho)}{d\rho}+
\left(E_B{\rho^2}-\frac{n^2}{B^2}\right)\psi(\rho)=0.
\label{15}
\end{equation}
Equation (\ref{15}) can be recast in a more suitable form with the position $q=b\rho$, where $b=\sqrt{E_B}$ ($E_B\geq 0$); 
hence we can write
\begin{equation}
{q}^2\frac{d^2\psi(q)}{d{q}^2}+q\frac{d\psi(q)}{dq}+
\left(q^2-\frac{n^2}{B^2}\right)\psi(q)=0.
\label{16}
\end{equation}
For $B=1$, (\ref{16}) is nothing else but Bessel's equation whose solutions can be expressed in terms of 
Bessel functions $J_n(s)$ with $n\in\mathbb{N}$. In our case, the presence of the topological defect makes the 
usual solutions $J_n$ no more suitable. Instead, we need the Bessel's functions with non integer index $\nu$,
$\nu=n/B$.
\section{Mathematical treatment of Hamiltonians with two attractive delta interactions}

In this section we exploit the technique used in \cite{1} in order to depict an attractive delta interaction on the plain placed on the z-axis, rather than two attractive delta interactions in a 1D framework treated in \cite{1}.
\\
To start with, note that equation (\ref{8}) it looks formally identical to the one in \cite{1}. It is worth noticing that in the present paper the support of our two attractive $\delta $-interactions is only on the plane $z=\pm z_{0}$.\\
Giving that our Hamiltonian is separable, we can explicit the $1$D Hamiltonian studied in \cite{1} to deal with a $3$D 
model involving a $\delta$-interaction supported on the planes $z=\pm z_0$. For the sake of completeness, we repeat the calculations present 
in \cite{1}, but adapted to our $3$D model.

Mimicking thus the procedure presented in \cite{1}, the Hamiltonian operator (\ref{8}),
with the dependence by the z-coordinate
as the limit in the norm resolvent sense, can be expressed as

\begin{equation}
	H^{\left\{\epsilon\right\}}_{\left\{\lambda,z_{0}\right\}}=-\frac{d^{2}}{dz^{2}}-\frac{\lambda}{\epsilon}V(\frac{z+z_{0}}{\epsilon})-\frac{\lambda}{\epsilon}V(\frac{z-z_{0}}{\epsilon}).
	\label{17}
\end{equation}
Note that the present paper from thereafter is written using the momentum space ${\bf p}$, a Fourier-transformed quantity $A(z)$ will be indicated by $\hat{A}(p)$. The symmetry condition is imposed on the potential $V(z)=V(-z)$ in order to guarantee that $\hat{V}(p)$ is real.
Moreover, we consider bound states with $E_z<0$ that can be obtained in terms of the resolvent of 
$H^{\left\{\epsilon\right\}}_{\left\{\lambda,z_{0}\right\}}$. As shown in \cite{1}, to this purpose it is sufficient to consider the operator with integral kernel given by
\begin{eqnarray}
	& & B_{\{ \lambda ,z_{0} \} }^{(\varepsilon )} (p_z,p_z';E_z)=\label{18}\\
	& & =(p_z^{2} -E_z)^{-1/2} \lambda \mathop{V}\limits^{\wedge } (\varepsilon (p_z-p_z'))\frac{e^{i(p_z-p_z')z_{0} } +
		e^{-i(p_z-p_z')z_{0} } }{(2\pi )^{1/2} } (p_z'^{2} -E_z)^{-1/2},\nonumber 
\end{eqnarray}
with $E_z<0$.
Since $B_{\{ \lambda ,z_{0} \} }^{(\varepsilon )}(E_z)$ is positive trace class operator, it can be expressed as the sum of two operators, which are unitarily equivalent to the positive trace class operator, whose integral kernel (see \cite{1} and references therein) is given by
\begin{equation}
	(p_z^{2} -E_z)^{-1/2} \frac{\lambda \mathop{V}\limits^{\wedge } (\varepsilon (p_z-p_z'))}{(2\pi )^{1/2} } (p_z'^{2} -E_z)^{-1/2} ,\forall E_z<0, 
	\label{19}
\end{equation}
and for the trace of $B_{\{ \lambda ,z_{0} \} }^{(\varepsilon )}(E_z)$, we get
\begin{eqnarray}	
	& & \left\| B_{\{ \lambda ,z_{0} \} }^{(\varepsilon )} (E)\right\| _{1} =\frac{2\lambda \mathop{V}\limits^{\wedge } (0)}{(2\pi )^{1/2} } \cdot \int _{-\infty }^{\infty }dp_z (p_z^{2} -E_z)^{-1} =\nonumber\\
	& & \frac{\lambda }{\left|E_z\right|^{1/2} } \int _{-\infty }^{\infty }V(z)dz =\frac{\lambda }{\left|E_z\right|^{1/2} } =\left\| B\right. _{\{ \lambda ,z_{0} \} } \left. (E_z)\right\| _{1},    
	\label{20}
\end{eqnarray}
where $B_{\{ \lambda ,z_{0} \} } (E_z)$ refers to the part of the Hamiltonian (\ref{4}) depending only on the $z$ coordinate. 
Moreover, for
$\varepsilon \to 0_{+} $\textit{,} $B_{\{ \lambda ,z_{0} \} }^{(\varepsilon )} (E_z)$ converges weakly to $B_{\{ \lambda ,z_{0} \} } (E_z)$
and the convergence is obtained in the Banach space of trace class operators.\\
By following the treatment present in \cite{1}, the limiting rank-two operator has the integral kernel, that in Dirac notation can be written as
\begin{eqnarray}
	& & B_{\{ \lambda ,z_{0} \} } (E_z)=\frac{\lambda }{\pi } {\left| \psi _{\{ z_{0,} ,E_z\} }^{(0)}  \right\rangle} {\left\langle \psi _{\{ z_{0,} ,E_z\} }^{(0)}  \right|} +\frac{\lambda }{\pi } {\left| \psi _{\{ z_{0,} ,E_z\} }^{(1)}  \right\rangle} {\left\langle \psi _{\{ z_{0,} ,E_z\} }^{(1)}  \right|} =\nonumber\\
	& & =\frac{\lambda }{\pi } \left\| \psi \right. \left. _{\{ z_{0,} ,E_z\} }^{(0)} \right\| _{2}^{2} P_{\{ z_{0,} ,E_z\} }^{(0)} \frac{\lambda }{\pi } \left\| \psi \right. \left. _{\{ z_{0,} ,E_z\} }^{1} \right\| _{2}^{2} P_{\{ z_{0,} ,E_z\} }^{1},
	\label{21}
\end{eqnarray}
where $P_{\{ z_{0,} ,E_z\} }^{(0)}$ and $P_{\{ z_{0,} ,E_z\} }^{(1)}$ are the respective orthogonal projectors, with
$\psi _{\{ z_{0,} ,E_z\} }^{(0)} =\frac{\cos (z_{0} \cdot p_z)}{(p_z^{2} -E_z)^{1/2} }$ the even function, while
$\psi _{\{ z_{0} ,E_z\} }^{(1)} =\frac{\sin (z_{0} \cdot p_z)}{(p_z^{2} -E_z)^{1/2} }$ is the odd one, with obviously
$(\psi _{\{ z_{0} ,E_z\} }^{(0)} ,\psi _{\{ z_{0} ,E_z\} }^{(1)} )=0$. Hence, for the two eigenvalues of the operator
$B_{\{ \lambda ,z_{0} \} }(E_z)$ we get
\begin{eqnarray}
	& &\beta _{\{ \lambda ,z_{0} \} }^{(0)} (E_z)=\label{22}\\
	& & =\frac{\lambda }{2\pi } \int _{-\infty }^{+\infty }dp_z \frac{1+\cos (2z_{0} \cdot p_z)}{p_z^{2} +\left|E_z\right|} =\frac{\lambda }{2\left|E_z\right|^{1/2} } \left(1+e^{-2z_{0} \left|E_z\right|^{1/2} } \right),\nonumber\\ 
	& & \beta _{\{ \lambda ,z_{0} \} }^{(1)} (E_z)=\label{23}\\ 
	& & =\frac{\lambda }{2\pi } \int _{-\infty }^{+\infty }dp_z \frac{1-\cos (2z_{0} \cdot p_z)}{p_z^{2} +\left|E_z\right|} =\frac{\lambda }{2\left|E_z\right|^{1/2} } \left(1-e^{-2z_{0} \left|E_z\right|^{1/2} } \right).\nonumber
\end{eqnarray}
In order to obtain the eigenvalues of $H_{\{ \lambda ,z_{0} \}}$, as the poles of its resolvent, as shown in \cite{1},
we need to express the operator $[B_{\{ \lambda ,z_{0} \} } (E_z)]^{n}$, with $n\in\mathbb{N}$, together with the Neumann series of $[1-B_{\{ \lambda ,z_{0} \} } (E_z)]^{-1}$, with the following equality
\begin{equation}
	[B_{\{ \lambda ,z_{0} \} } (E_z)]^{n} =[\beta _{\{ \lambda ,z_{0} \} }^{(0)} (E_z)]^{n} P_{\{ z_{0,} ,E_z\} }^{(0)} +[\beta _{\{ \lambda ,z_{0} \} }^{(1)} (E_z)]^{n} P_{\{ z_{0,} ,E_z\} }^{(1)}.
	\label{24} 
\end{equation}
Hence, the resolvent of $H_{\{ \lambda ,z_{0} \}}$
in the $p_z-$space is given by:
\begin{eqnarray}
	& &\left[H_{\{ \lambda ,z_{0} \} } -E_z\right]^{-1}=(p_z^{2} -E_z)^{-1}+\label{25}\\
	& &+\frac{\frac{\lambda }{\pi } }{1-\frac{\lambda }{\pi } \left\| \psi \right. \left. _{\{ z_{0,} ,E_z\} }^{(0)} \right\| _{2}^{2} } {\left| (p_z^{2} -E_z)^{-1/2} \psi _{\{ z_{0} ,E_z\} }^{(0)}  \right\rangle} {\left\langle (p_z^{2} -E_z)^{-1/2} \psi _{\{ z_{0} ,E_z\} }^{(0)} \right|} +\nonumber\\
	& &+\frac{\frac{\lambda }{\pi } }{1-\frac{\lambda }{\pi } \left\| \psi \right. \left. _{\{ z_{0,} ,E_z\} }^{(1)} \right\| _{2}^{2} } {\left| (p_z^{2} -E_z)^{-1/2} \psi _{\{ z_{0} ,E_z\} }^{(1)}  \right\rangle} {\left\langle (p_z^{2} -E_z)^{-1/2} \psi _{\{ z_{0} ,E_z\} }^{(1)}  \right|}.\nonumber 
\end{eqnarray}
After performing the Fourier antitrasform of the $p_z-$space and thus expressing the resolvent (\ref{25})
in the $z-$space, for the ground state energy, the poles of the resolvent (\ref{25}) in the $z-$space, thanks to the eigenvalue
(\ref{22}), we get
\begin{equation}
	\frac{2\left|E_z^{(0)}\right|^{1/2} }{1+e^{-2z_{0} \left|E_z^{(0)}\right|^{1/2} } } =\lambda.
	\label{26}
\end{equation}
In a similar manner, by using the eigenvalue (\ref{23}), for the energy of the other possible bound state we get
\begin{equation}
	\frac{2\left|E_z^{(1)}\right|^{1/2} }{1-e^{-2z_{0} \left|E_z^{(1)}\right|^{1/2} } } =\lambda.
	\label{27} 
\end{equation} 
The implicit expressions for $E_z^{(0)}$ and $E_z^{(1)}$ will be analyzed in the next section.

\section{Spectrum of point-mass particles perturbed by two attractive delta interactions}
 
In obtaining (\ref{26}) and (\ref{27}), we used the mass normalization $M=1/2$ together with  $G=c=\hbar=1$. 
By using the transformations
\begin{equation}
\lambda\rightarrow \frac{2M\overline{\lambda}}{{\hbar}^2},\;\;\;\;E_z\rightarrow\frac{2M{\overline{E}}_z}{{\hbar}^2},	
\label{28}
\end{equation}
(\ref{26}) and (\ref{27}) become respectively:
\begin{eqnarray}
& & \frac{\hbar\left|2M{\overline{E}}_z^{(0)}\right|^{1/2} }{1+e^{-\frac{2z_{0}}{\hbar}
	\left|2M{\overline{E}}_z^{(0)}\right|^{1/2} } } =M\overline{\lambda},\label{29}\\
& & \frac{\hbar\left|2M{\overline{E}}_z^{(1)}\right|^{1/2} }{1-e^{-\frac{2z_{0}}{\hbar}
			\left|2M{\overline{E}}_z^{(1)}\right|^{1/2} } }=M\overline{\lambda}.\label{30}
\end{eqnarray}
A first study of the equations (\ref{29})-(\ref{30}) can be performed using the same technique present in \cite{1}.
After introducing the adimensional variable $\xi$ with 
$\xi=\frac{z_0}{\hbar}\left|2M{\overline{E}}_z^{(0)}\right|^{1/2}$, we obtain the equivalent equation
\begin{equation}
F(\xi)=\frac{\xi }{1+e^{-2\xi } } = \frac{z_0 M\overline{\lambda}}{\hbar^2}. 
\label{31}
\end{equation}
Since $\frac{dF}{d\xi}>0,\;\forall\xi\geq 0$, we can invert (\ref{31}) with inverse $F^{-1}(\xi),\;\forall\xi\geq 0$.
In particular, we are interested in the limiting case with $z_0\rightarrow 0^+$ and $z_0\rightarrow +\infty$.
In the former case, we have
\begin{equation}
{\overline{E}}_z^{(0)}\rightarrow -\frac{2M{\overline{\lambda}}^2}{\hbar^2},\;\;\;for\;\;\;z_0\rightarrow 0^+,
\label{32}
\end{equation} 
while in the latter case we get
\begin{equation}
{\overline{E}}_z^{(0)}\rightarrow -\frac{M{\overline{\lambda}}^2}{2\hbar^2},\;\;\;for\;\;\;z_0\rightarrow +\infty.
\label{33}
\end{equation}
Moreover, the  derivative $\frac{d{\overline{E}}_z^{(0)}}{dz_0}$ can be easily computed by using implicit differentiation with the result
$\frac{d{\overline{E}}_z^{(0)}}{dz_0}>0$. Hence, ${\overline{E}}_z^{(0)}$ is an increasing function of $z_0$ for any fixed
$\overline{\lambda}$ and $M$.\\ 
Summarizing, we have 
${\overline{E}}_z^{(0)}\in(-\frac{2M{\overline{\lambda}}^2}{\hbar^2}, -\frac{M{\overline{\lambda}}^2}{2\hbar^2})$.

A similar study can be carried out for equation (\ref{30}). An important difference with respect to (\ref{29}) is that 
the counterpart of (\ref{31}), namely
\begin{equation}
G(\xi)=\frac{\xi }{1-e^{-2\xi } } = \frac{z_0 M\overline{\lambda}}{\hbar^2}, 
\label{34}
\end{equation}
has a removable singularity  for ${\overline{E}}_z^{( 1)}\rightarrow 0$ and (see \cite{1} for more details)
the inverse function $G^{-1}(\xi)$ is defined only on $[1,+\infty)$. Thus, the following existence condition,
present in \cite{1} with $M=1/2$,
\begin{equation}
2Mz_0\overline{\lambda}>\hbar^2
\label{35}
\end{equation}
must be satisfied. We remind the reader that such a condition coincides with (2.1.31) in section II.2.1 in \cite{B} ensuring the existence of the zero energy resonance \footnote{The latter is equivalent to the existence of the coupling constant threshold.}, in the special case of two identical attractive delta interactions symmetrically situated around the origin. It is interesting to point out that, 
for a given choice of the parameters $\{\overline{\lambda},z_0\}$, 
the existence condition 
(\ref{35}) imposes a minimal mass in order to be fulfiled. This means that particles with mass violating the (\ref{35}) cannot be in the excited states, hence they can only live in the ground state.
In a similar manner as for (\ref{32}) and (\ref{33}), we obtain
${\overline{E}}_z^{(1)}\rightarrow 0$ for $z_0\rightarrow 0^+$, and
${\overline{E}}_z^{(1)}\rightarrow -\frac{M{\overline{\lambda}}^2}{2\hbar^2}$ for $z_0\rightarrow +\infty$.
Due to (\ref{35}), the bound state energy in the limit for
$z_0\rightarrow+\infty$, can be obtained at a given finite value for $\overline{\lambda}$ only for massless particles with
$M\sim 1/z_0$.
Moreover, we have 
$\frac{d{\overline{E}}_z^{(1)}}{dz_0}<0$ and ${\overline{E}}_z^{(1)}$ is a 
decreasing function of $z_0$.\\
Summarizing, for the excited state we have ${\overline{E}}_z^{(1)}\in(-\frac{M{\overline{\lambda}}^2}{2\hbar^2},0)$.
For $z_0\rightarrow +\infty$ we obtain ${\overline{E}}_z^{(1)}\rightarrow {\overline{E}}_z^{(0)}$ with
${\overline{E}}_z^{(1)}>{\overline{E}}_z^{(0)}$.

\section{A physical application: double layer graphene}

In this section we apply the previous mathematical tools in order to provide a description of the double layer graphene (reduced to a single layer in the limit of $z_{0}\rightarrow0$) with a topological defect.
To be more precise the two layers placed at $z=\pm z_{0}$ represent the two graphene layers, while the cosmic string, with negative string tension (see \cite{21,21b,21c}), depicts the topological defect. Moreover, it is worth pointing out that in the usual graphene modeling, the double layers are to be considered very close to each other; this implies that $|z_{0}|<<1$. This kind of model is similar to the one present in \cite{21}, where the topological defect of the single graphene layer is given by the cosmic string. In a real conductor the particle is trapped within the cross section of the cylinder, hence Dirichlet boundary conditions must be imposed on the radius $0<r\leq R$, where $R$ is the radius of the infinitely long cylinder, with the two deltas placed at at $z=\pm z_{0}$. A fundamental feature, in order to modelize graphene as a conductor is the presence of bound states with $E<0$ (holes), together with ones of positive energy $E>0$ (electrons in a conductor). In practice, bound states with $E>0$ are representing electrons in the conduction band, while bound states with $E<0$ depict the dynamics of holes.
\\
To start with, in order to get the total spectrum of the trapped particle, we must study the solutions of (\ref{16}).
The aforementioned equation is nothing else but the one giving the Bessel's functions $\psi(q)\sim J_{\nu}(q)$
of first kind with non-integer order $\nu=n/B$. The expression for the solution of (\ref{16}) is given in terms of the 
series expansion
\begin{equation}
J_{\nu}(q)={\left(\frac{q}{2}\right)}^{\nu}\sum_{j=0}^{\infty}
\frac{{(-1)}^j}{j!\;\Gamma(j+\nu+1)}{\left(\frac{q}{2}\right)}^{2j},
\label{36}
\end{equation} 
with $\Gamma(s)=\int_0^{\infty} e^{-t}t^{s-1}dt$. For $\nu\notin\mathbb{N}$, the functions $J_{\nu}(q)$ and 
$J_{-\nu}(q)$ are linearly independent. We choose the regular solutions $J_{\nu}(q)$ with $\nu=n/B\geq 0$ and hence we 
set $n\in\mathbb{N}$. In order to impose the Dirichelet boundary condition 
$\psi(\rho={\rho}_0=R)=0\rightarrow J_{\nu}({\rho}_0)=0$,
we need an explicit expression for $J_{\nu}(q)$. From (\ref{36}), the following expansions for $q<<1$ and 
$q>>1$, with $\nu\geq 0$ can be obtained \cite{26}:
\begin{eqnarray}
& & J_{\nu}(q)\rightarrow \frac{1}{\Gamma(\nu+1)}{\left(\frac{q}{2}\right)}^2,\;\;\;as\;\;\;q<<1,\label{37}\\
& & J_{\nu}(q)\rightarrow \sqrt{\frac{2}{\pi q}}\cos\left(q-\frac{\pi\nu}{2}-\frac{\pi}{4}\right),\;\;\;as\;\;\;q>>1. \label{38}
\end{eqnarray}
The transition from the behavior for small $q$ and large $q$ does happen when $q\simeq\nu$.
In practice, (\ref{38}) is a sufficient approximation also for $q\geq\nu$. As a consequence of these results, we can obtain 
suitable approximations for the zeros $q_{\nu m}$ of the Bessel's functions with
$J_{\nu}(q_{\nu m})=0$ and $m\in\mathbb{N}$. In fact, for any $\nu\leq 2$, zeros are approximatively separated 
by $\pi$ (see \cite{26}):
\begin{equation}
{\overline{E}}_{Bnm}\simeq \frac{{\hbar^2\left({\alpha}_{\nu}+m\pi\right)}^2}{2 M R^2},\;\;\;m\in\mathbb{N}.
\label{39}
\end{equation}
For example, choosing ${\alpha}_{0}=2.405, {\alpha}_{1}=3.832, {\alpha}_{2}=5.136$, 
${\alpha}_{\nu}\in[2.405,5.136]$ for $\nu\in[0,2]$.\\ 
For $\nu>2$, a suitable approximation for the roots of the
equation $J_{\nu}(q(R))=0$ is given by:
\begin{equation}
q_{nm}=\frac{\pi n}{2B}+\pi m+\frac{3\pi}{4},
\label{40}
\end{equation}
with
\begin{equation}
{\overline{E}}_{Bnm}={\overline{E}}_{\nu m}\simeq
\frac{\pi^2\hbar^2}{2M R^2}{\left(\frac{n}{2B}+m+\frac{3}{4}\right)}^2.
\label{41}
\end{equation}
Note that, for $B=1$ or $\nu\in\mathbb{N}$, the Bessel's expressions with integer order are regained, while for
$\nu=n+1/2$ we obtain the spherical Bessel's functions.\\
In any case, (\ref{41}) gives rise to an acceptable approximation also for the quantum states 
(\ref{39}), with $\nu=\{0,1,2\}$ and, as a consequence, the expression (\ref{41}) provides a sufficient expression for our purposes.
Hence, as a consequence of the results on the admissible values for
${\overline{E}}_z^{(0)}$ and ${\overline{E}}_z^{(1)}$, we get:
\begin{equation}
{\overline{E_t}}_{nm}={\overline{E}}_{\nu m}-\frac{H(z_0,\overline{\lambda}) M{\overline{\lambda}}^2}{\hbar^2},
\label{42}
\end{equation}
with $H(z_0,\overline{\lambda})\in (1/2,2)$ for the ground state energy ${\overline{E}}_z^{(0)}$ and 
$H(z_0,\overline{\lambda})\in (0,1/2)$ for the excited state energy ${\overline{E}}_z^{(1)}$. The spectrum given by (\ref{42}) is obviously discrete with quantum numbers $\{n,m\}\in\mathbb{N}$ and physical parameters $\{R, M,\overline{\lambda}, z_0,B\}$.\\
The bound states are the ones with ${\overline{E_t}}_{nm}<0$. \\
It is physically relevant to study possible configurations 
separating bound states with ${\overline{E_t}}_{nm}<0$ from the ones with ${\overline{E_t}}_{nm}>0$.
To this purpose, note that, at any fixed quantum state with quantum numbers $\{n,m\}$ and with any choice of the parameters
$\{M,\overline{\lambda}, z_0,B\}$, there exists a radius ${\overline{R}}_{nm}$ for the trapping cylinder such that for 
$R={\overline{R}}_{nm}$ we obtain 
${\overline{E_t}}_{nm}=0$:
\begin{equation}
{\overline{R}}_{nm}=\frac{\pi\hbar^2}{M\overline{\lambda}\sqrt{2H(z_0,\overline{\lambda})}}
\left(\frac{n}{2B}+m+\frac{3}{4}\right).
\label{43}
\end{equation}
Equality (\ref{43}) shows interesting physical consequences. By fixing the physical parameters
$\{M,\overline{\lambda},z_0,B\}$ and the integers $\{n,m\}$, with $\{n,m\}=\{\overline{n},\overline{m}\}$, it is always possible to 'build' the trapping cylinder with radius 
${\overline{R}}_{nm}={\overline{R}}_{\overline{n}\overline{m}}$. From (\ref{42}) we deduce that particles in quantum states, with quantum numbers $\{n,m\}$, satisfying the inequality
\begin{equation}
\left(\frac{n}{2B}+m\right) < \left(\frac{\overline{n}}{2B}+\overline{m}\right),
\label{44}
\end{equation}
are in bound states with ${\overline{E_t}}_{nm}<0$, while the ones satisfying the opposite inequality are in states with
${\overline{E_t}}_{nm}>0$ and finally, the ones with quantum numbers $\{n,m\}=\{\overline{n},\overline{m}\}$, have 
${\overline{E_t}}_{nm}=0$. In practice, apart from the case with 
$\{\overline{n},\overline{m}\}=\{0,0\}$, where ${\overline{E_t}}_{nm}>0,\;\forall\{n,m\}\neq\{0,0\}$,
in a more general framework, with $\{\overline{n},\overline{m}\}\neq\{0,0\}$, we can observe states with 
${\overline{E_t}}_{nm}\geq0$ and bound states with ${\overline{E_t}}_{nm}<0$.\\
Concerning the role of $B$, depicting the topological defect, 
it is worth pointing out that from (\ref{43}), by fixing the all other physical parameters, 
${\overline{R}}_{nm}$ is a monotonic decreasing function of $B$.
Hence, for $B\in(0,1)$, suitable with positive string-density tension in a general relativistic context, 
the value of
${\overline{R}}_{nm}$ is greater than the one in the case with $B\in(1,+\infty)$, 
with a negative string-density tension, suitable in the framework of nanophysics, electrons in a finite metal, semiconductors or graphene
modeling. This fact is in agreement with physical intuition, since we expect that a negative string-density tension
favors bound states with ${\overline{E_t}}_{nm}<0$.

\section{Conclusions and final remarks}

In this paper we studied the discrete spectrum arising 
for a mass particle trapped in an infinitely long cylinder with two attractive delta-interactions along the planes 
$z=\pm z_0$, with a topological defect (cosmic string) along the $z-$axis. In this framework, the paper provides a first attempt to apply the technique in \cite{1} to a curved background with the Laplace-Beltrami operator. 
The physical effects shown by a cosmic string are very similar to the well-known Aharonov-Bohm effect 
\cite{13} in quantum mechanics, since the string determines a deviation on the trajectory of a particle, despite the locally flat character of the manifold.\\
With the value of the deficit angle $B\in(0,1)$, we have a string with a positive energy density tension; hence, this kind of modeling could be of interest in the context of general relativity and cosmology, where topological defects are expected to play a role in
the primordial history of our universe (primordial inflation). Conversely, for $B\in(1,+\infty)$, we observe an energy momentum 
tensor $T_{\mu\nu}$, with a negative energy-density tension on the $z-$axis. A negative tension for the string can be used in the 
framework of \cite{21}. In this article we have focused our attention on a model representing a mass particle trapped inside an infinitely long cylinder with a topological defect depicted by a string. The resulting Hamiltonian has been perturbed by two attractive delta potentials lying on the parallel planes $z=\pm z_0$, so that it is also separable with respect to the cylindrical coordinates
$\{\rho,z,\phi\}$. Concerning the treatment of the part of the Hamiltonian depending on the $z$ coordinate, we used the machinery developed in \cite{1}, to write down the spectrum for the two bound states, due to the two attractive delta perturbations at $z=\pm z_0$. For the spectrum resulting from the Hamiltonian depending on
the coordinates ${\rho,z}$, we have exploited the Bessel's functions $J_{\nu}(q)$,
with non-integer order, further we have obtained a suitable approximating formula for the zeros of $J_{\nu}(q)$. The joint spectrum has thus been studied in section 6.
Further, an interesting consequence of our application is the general presence of a mixing 
between discrete states with positive total energy, together with bound states and ones with zero energy.
Infact, as explicitely shown in section 6, such a procedure is used in the framework of the physics of graphene.
Specifically, the model presented in this article could provide a way to depict the double-layer graphene, given the great interest it has drawn in the last two decades.
Moreover, we can consider particles with a given mass $M$. Following the result of section 4, and in particular due to
(\ref{43}), by fixing the radius ${\overline{R}}_{nm}$ to 
${\overline{R}}_{nm}={\overline{R}}_{00}$, and by physical transformations, we are able to recast bosons in the state 
$\{n,m\}=\{0,0\}$. In this case, the total energy is zero. This situation
can be observed also in a generic excited state provided by:
\begin{equation}
\left(\frac{n}{2B}+m\right) = \left(\frac{\overline{n}}{2B}+\overline{m}\right).
\label{45}
\end{equation}
In some sense, this configuration could represent an analogous realization of the Bose-Einstein condensation (BEC), also for excited states as proposed in \cite{27}, by using a generalization of the 
Gibbs distribution \cite{28}.

It is worth to notice that, within our modeling in the limit for $R\rightarrow 0$, a particle is confined in a 3D quantum dot. However, in this case is to be studied the appropriate behavior of the Bessel function; this could be matter for a further investigations.

As a final consideration, our model could be also of interest to mimic, in some sense, a black hole. In this regard, note that, the
part of the spectrum due to the hard wall at ${\rho}=R$, is very similar to the one obtained in \cite{23} depicting gravitons 
inside the black hole.  

\section*{Acknowledgements} 

We wish to thank the anonimous referee for his/her precious suggestions in order to greatly improve the content and the presentation of this paper.
\\
We also wish to thank Prof. S. Albeverio for his constant encouragement and
stimulating mentorship. Financial support is gratefully acknowledged by S.
Fassari to the Spanish Junta de Castilla y Le\'on and FEDER (Project
VA057U16) and MINECO (Project MTM201457129-C2-1-P). S. Fassari wishes
to thank the entire staff at ''Departamento de Fi­sica Teorica, Atomica y
Optica, Universidad de Valladolid'', for their warm hospitality
throughout his stay.

\end{document}